# Độ không đảm bảo đo trong phép đo xác định độ không phẳng bàn máp

*Lê Hữu Thắng – Phòng Đo lường Độ dài - Viện Đo lường Việt Nam*

**Tóm tắt**: Độ không phẳng (ĐKP) của bàn máp là thông số quan trọng và có rất nhiều qui trình tính cũng như các nghiên cứu xác định các yếu tố ảnh hưởng tới độ chính xác của phép đo này. Tuy nhiên các nghiên cứu đó nằm rải rác khó tập hợp, thêm vào đó không phải lúc nào các báo cáo này cũng được viết hài hoà với Hướng dẫn biểu diễn độ không đảm bảo đo (GUM) của Tổ chức đo lường quốc tế BIPM. Hơn thế nữa cách viết mô hình toán miêu tả phép đo ĐKP một cách tường minh cũng không thấy xuất hiện. Để lấp đi các thiếu hụt nói tới ở trên, chúng tôi đã tiến hành tập hợp lại các yếu tố ảnh hưởng tới độ chính xác của phép đo xác định ĐKP một cách hệ thống, trình bày tường minh phương trình miêu tả phép đo xác định biên dạng của bề mặt bàn máp và đề xuất phương trình miêu tả việc xác định ĐKP dựa theo định nghĩa của ISO. Xuất phát từ các phương trình này, độ không đảm bảo đo (ĐKĐBĐ) của ĐKP đo được sẽ được xác định hài hoà với nội dung của GUM. Ví dụ số minh hoạ cũng được trình bày.

**Abstract**: Flatness of a plate is a parameter has been put under consideration for long time. Factors influencing the accuracy of this parameter have been recognized and examined carefully but placed scatterringly. Beside that those reports have not been always in harmonization with Guide for expression of uncertainty measurement (GUM). Furthermore, mathematical equations describing clearly the flatness measurement have not been seen in those reports also. We have collected those influencing factors for systematic reference purpose, re-written the equation describing the profile measurement of the plate topography, and proposed an equation for flatness determination. An illustrative numerical example will be also shown.

**I. Giới thiệu**

Độ không phẳng (ĐKP) của bàn máp là thông số quan trọng và có rất nhiều qui trình tính cũng như các nghiên cứu về các yếu tố ảnh hưởng tới độ chính xác của phép đo này. Tuy nhiên các nghiên cứu đó không phải lúc nào cũng được trình bày hài hoà với Hướng dẫn biểu diễn độ không đảm bảo đo (GUM) của Tổ chức đo lường quốc tế BIPM. Hơn thế nữa mô hình toán miêu tả phép đo ĐKP một cách tường minh cũng không thấy xuất hiện. Để lấp đi các thiếu hụt nói tới ở trên, chúng tôi đã tiến hành tập hợp lại các yếu tố ảnh hưởng tới độ chính xác của phép đo xác định ĐKP một cách hệ thống, viết lại phương trình miêu tả phép đo xác định biên dạng của mặt phẳng bàn máp và đề xuất phương trình miêu tả việc xác định ĐKP. Xuất phát từ các phương trình này, ĐKĐBĐ của ĐKP đo được sẽ được xác định hài hoà với các hướng dẫn của GUM.

Có nhiều phương pháp khác nhau xác định ĐKP của bàn máp bằng các phương tiện đo khác nhau [2 - 5]. Trong các phương pháp này người ta xác định cao độ của các điểm tại từng vị trí xác định trước dọc theo các đường thẳng theo sơ đồ lưới điểm đo trên toàn bộ mặt phẳng đang được quan tâm. Sơ đồ các lưới điểm này có thể không giống nhau tuỳ thuộc vào phương pháp cụ thể. Ở giai đoạn tính toán thông số ĐKP, người ta tìm cách xác định mặt phẳng chuẩn là mặt phẳng sao cho tồn tại hai mặt phẳng song song với nó mà bao toàn bộ các điểm số liệu đo được. Thực tế ta sẽ tìm được nhiều hơn một mặt phẳng như vậy. Theo định nghĩa của ISO, khoảng cách giữa hai mặt phẳng song song này sẽ là ĐKP cần xác định nếu nó thoả mãn điều kiện là đoạn thẳng ngắn nhất.

Một trong các sơ đồ đo đơn giản nhất cho việc xác định ĐKP của bàn máp là sơ đồ được trình bày trong hình 1 dưới đây [6], trong đó nivô có thể được sử dụng làm thiết bị đo. Các điểm đo được đánh chữ từ 1 – 9 dọc theo các đường tương ứng, tạo nên lưới các điểm đo. Như vậy hệ thống lưới các điểm này với cao độ đã được xác định cho ta hình dạng của biên dạng của mặt phẳng đang được quan tâm. Có nhiều yếu tố ảnh hưởng tới độ chính xác của việc xác định đường biên dạng này và chúng sẽ được phân tích ở phần II sau đây.

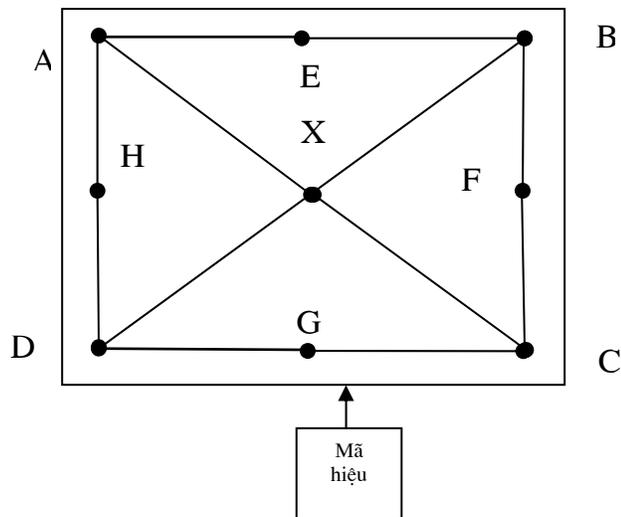

*Hình 1. Sơ đồ xác định độ không phẳng của bàn máp. Các số từ A – X là tên của các điểm đo. Ghi chú: các kết quả đo tại cùng một điểm trên hai đường khác nhau có thể có gí trị đo không giống nhau do việc lấy số liệu trên hai đường này là độc lập với nhau.*

## II. Mô hình toán và độ không đảm bảo đo thành phần trong việc xác định biên dạng của bàn máp

Để ước tính ĐKĐBĐ, theo GUM trước tiên ta cần nhận biết và tìm hiểu các yếu tố ảnh hưởng tới độ chính xác của phép đo đang được xem xét, sau đó xác định dạng của hàm mật độ phân bố xác suất tương ứng của chúng. ĐKĐBĐ chuẩn của các yếu tố này sau đó sẽ được ước tính. Trong trường hợp phép đo xác định biên dạng của mặt bàn máp, ĐKĐBĐ kết hợp của phép đo này được tính thông qua mô hình toán sau:

$$\theta = \theta_s + \theta_x + \theta_t + \theta_{cd} + \theta_{td} + \theta_{ss} + \theta_{da} \quad (1)$$

Trong phương trình này, các yếu tố ảnh hưởng tới độ chính xác của phép đo xác định góc nghiêng $\theta$ (mm/m) hay tương ứng là cao độ $h$ (mm) của điểm đo bị ảnh hưởng bởi tám yếu tố thành phần như trong phương trình (1). Ý nghĩa và hàm mật độ phân bố xác suất của từng yếu tố ảnh hưởng này sẽ được giới thiệu trong bảng 1 và các phần ngay sau đó.

*Bảng 1. Các yếu tố ảnh hưởng tới độ chính xác của phép đo hiệu chuẩn độ không phẳng bàn máp và các hàm phân bố mật độ xác suất tương ứng*

| Tên yếu tố ảnh hưởng | Ký hiệu | Dạng hàm phân bố |
|---|---|---|
| 1. Thiết bị đo | $\theta_s$ | Gauss |
| 2. Độ cong trái đất | $\theta_{td}$ | Chữ nhật |
| 3. Độ lặp lại | $\theta_x$ | t (student) |
| 4. Gradient nhiệt | $\theta_t$ | Chữ nhật |
| 5. Hệ chân đỡ | $\theta_{cd}$ | Chữ nhật |
| 6. Khớp điểm đo | $\theta_{ss}$ | Chữ nhật |
| 7. Độ ẩm | $\theta_{da}$ | Chữ nhật |

*1. Thiết bị đo*

Phương tiện đo (PTĐ) được dùng trong phép đo xác định ĐKP thường là nivô hoặc ống tự chuẩn trực. Thông thường với mỗi một PTĐ luôn có đi kèm một giấy chứng nhận hiệu chuẩn do nhà sản xuất thiết bị cung cấp, trong đó có ghi độ chính xác và ĐKĐBĐ hiệu chuẩn thiết bị. ĐKĐBĐ này thường được viết dưới dạng: $U(\theta_s) = 2\,u(\theta_s), k = 2, P = 95\%..$

*2. Độ cong của trái đất*

Do trái đất có độ cong khác không nên mỗi phép đọc số liệu đo từ nivô đều mắc phải sai số đọc. Giá trị cực đại của sai số góc này so với điểm đo nằm cách nó một khoảng cách $l$ có dạng: $\theta_{td} = \dfrac{l}{8r}$ (2). ĐKĐBĐ tương ứng với thành phần sai số này sẽ là:

$u(\theta_{td}) = \dfrac{\theta_{td}}{2\sqrt{3}}$ (3). Ở đây $r$ – bán kính trái đất trung bình.

*3. Độ lặp lại*

Độ lặp lại được tính theo công thức quen thuộc: $u(\theta_x) = \dfrac{\sqrt{\sum_{i=1}^{N}(\theta_i - \bar{\theta}_i)}}{N(N-1)}$ (4), trong đó $N$ là số lần đo lặp, $i$ chỉ số lần đo thứ $i$ và giá trị trung bình số học $\bar{\theta}_i$.

*4. Gradient nhiệt*

Nếu nhiệt độ tại bề mặt trên và bề mặt dưới của bàn máp có chiều dày *d*, khác nhau không quá $\Delta T$ ($^0$C), thì góc cong vênh cực đại dọc theo đường thẳng *l* có dạng: $\theta_t = \dfrac{\alpha l^2 \Delta T}{8d}$ (5), và ĐKĐBĐ tương ứng sẽ có dạng: $u(\theta_t) = \dfrac{\theta_t}{2\sqrt{3}}$ (6).

*5. Hệ chân đỡ*

Theo khuyến cáo của các nghiên cứu trong tham khảo [13], ba chân đỡ của bàn máp nên đặt tại ba điểm sao cho chúng ở các vị trí đầu gần cạnh bàn nhất một khoảng cách 20% chiều dài của toàn cạnh bàn đó. Nếu điều kiện này được thực hiện thì ảnh hưởng do cách đặt chân đỡ lên độ phẳng sẽ là tối thiểu, tuy nhiên có giá trị khác không và bằng: $\theta_{cd} = gk\dfrac{\rho\omega^4(1-\nu^4)}{Ed^2 l}$ (7). ĐKĐBĐ tương ứng có dạng: $u(\theta_{cd}) = \dfrac{\theta_{cd}}{2\sqrt{3}}$ (8). Ở đây trong công thức (7) ta có gia tốc trọng trường *g*, hằng số *k*, ứng suất đàn hồi *E*, hệ số poisson $\nu$, các kích thước dài - rộng – cao *l – w – d* của bàn máp.

*6. Khớp điểm đo*

Tại các điểm giao của các đường lấy số liệu, các giá trị đo đọc được trên cùng một vị trí tại các lần đo khác nhau dọc theo các đường khác nhau có thể có giá trị khác nhau. Trị tuyệt đối của các sai khác này gọi là sai số đóng và ĐKĐBĐ tương ứng của nó có dạng: $u(\theta_{ss}) = \dfrac{\theta_{ss}}{2\sqrt{3}}$ (9). Ở đây giá trị $\theta_{ss}$ thường được chọn là giá trị cực đại trong số các giá trị sai số khớp điểm đo.

*7. Độ ẩm*

Các kết quả thực nghiệm cho thấy việc thay đổi độ ẩm không khí không gây ảnh hưởng tới việc xác định các cao độ của các điểm trên lưới đo. Tuy nhiên việc làm ướt mặt bàn bằng nước sẽ gây ra sự thay đổi các giá trị đo này tới tận 36 μm/m$^2$. ngay sau khi làm ướt mặt bàn. Tuy nhiên sự thay đổi này sẽ giảm về không với tốc độ rất chậm. Phải sau một vài tháng, sai lệch mới giảm về không [17 - 18].

**III. Thuật toán tính**

Trong các phương trình từ 1 đến 9, giá trị góc đo của các cao độ được sử dụng. Tuy nhiên ta có thể chuyển từ các giá trị góc thành thông số cao độ với thứ nguyên chiều dài bằng cách nhân chúng với chiều dài chân đo của nivô H = 150 mm. Cao độ $h = H\theta$ (10). Theo ISO, khoảng cách cực tiểu giữa hai mặt phẳng bao là ĐKP. Giả thiết ĐKĐBĐ của tất cả các điểm nằm trên hai mặt phẳng bao này là bằng nhau, nên mô hình toán của ĐKP sẽ phải là: ĐK$P = h_1 - h_2$ (11). Ở đây $h_1$ và $h_2$ là cao độ của hai điểm bất kỳ trên hai mặt phẳng bao sao cho đường thẳng đi qua hai điểm này vuông góc với cả hai mặt phẳng đó. Giả thiết ĐKĐBĐ của việc xác định cao độ của các điểm đo là như nhau, từ (11) ta có $u^2(ĐKP) = u^2(h_1) + u^2(h_2) = 2u^2(h_1) = 2u^2(h_2) = 2H^2 u^2(\theta)$ (12), hay $u(ĐKP) = \sqrt{2}H u(\theta)$ (13).

***Ví dụ số:***

Để minh hoạ cho cách tính ĐKĐBĐ trong phép đo xác định ĐKP của bàn máp, xét ví dụ với các số liệu thông thường như sau:

Bàn đá cấp II kích thước 250x250x150 mm.
Thiết bị đo: nivô số hiệu V01. TB3. 14, ĐKĐBĐ $u(\theta_s) = 0,001 mm/m, k = 2$
Độ chênh lệch nhiệt độ giữa mặt bàn trên và dưới là 0,4 $^0$C.
Hệ số dãn nở nhiệt tuyến tính α = 5,5x10$^{-6}$ K$^{-1}$; gia tốc trọng trường $g$ = 9,8 m s$^{-2}$; hệ số $k$ = 0,004; hệ số Poisson $v = 0,26$; ứng suất đàn hồi $E$ = 77x10$^9$ N K$^{-2}$; khối lượng riêng $\rho$ = 2750 kg m$^{-3}$ ; bán kính trái đất trung bình r = 6,37x10$^6$ m. Sơ đồ đo giống như trong hình 1, tuân theo hướng dẫn của ví dụ tương ứng trong JIS.
Thay các giá trị của các hằng số này và thực hiện các phép tính cần thiết ta có Bảng tổng kết ĐKĐBĐ (bảng số 2).

*Bảng 2. ĐKĐBĐ trong phép đo xác định ĐKP bàn máp với số liệu trên.*

| **Tên yếu tố ảnh hưởng** | **Ký hiệu** | **Dạng hàm phân bố** | **ĐKĐBĐ (mm/m)** | **Tỷ lệ %** |
|---|---|---|---|---|
| 1. Thiết bị đo | $\theta_s$ | Gauss | 5.00E-04 | 11.94 |
| 2. Độ cong trái đất | $\theta_{td}$ | Chữ nhật | 6.94E-06 | 0.00 |
| 3. Độ lặp lại | $\theta_x$ | t (student) | 1.00E-04 | 0.48 |
| 4. Gradient nhiệt | $\theta_t$ | Chữ nhật | 7.07E-04 | 23.88 |
| 5. Hệ chân đỡ | $\theta_{cd}$ | Chữ nhật | 1.55E-10 | 0.00 |
| 6. Khớp điểm đo | $\theta_{ss}$ | Chữ nhật | 1.15E-03 | 63.69 |
| 7. Độ ẩm | $\theta_{da}$ | Chữ nhật | - | - |

Từ số liệu tính trong bảng 2, áp dụng công thức số (13) cho việc tính ĐKĐBĐ cho *ĐKP* bàn máp ta có: u(ĐK*P*) = $\sqrt{2}$ *x* 0,015 *x* u($\theta$) = 0,0027 *mm*. ĐKĐBĐ mở rộng: U(ĐK*P*) = 2 u(ĐK*P*) = 0,0055 *mm*, *k* = 2 (*P* = 95%). Kết quả ĐKP tính theo cách miêu tả trong JIS đối với phép đo này là 0,013 mm.

**IV Tóm tắt**

Các yếu tố có thể ảnh hưởng tới độ chính xác của phép đo hiệu chuẩn *ĐKP* của bàn máp đã được tập hợp lại một cách tương đối đầy đủ. Mô hình toán miêu tả phép đo xác định biên dạng của mặt phẳng bàn máp đã được trình bày theo thông ước. Đặc biệt trong báo cáo này phương trình miêu tả việc xác định *ĐKP* đã được đề xuất. Từ đó, ĐKĐBĐ của ĐKP đo được sẽ được xác định hài hoà với các yêu cầu trong GUM.